# Spatially-resolved electronic and vibronic properties of single diamondoid molecules

YAYU WANG[1]*, EMMANOUIL KIOUPAKIS[1], XINGHUA LU[1], DANIEL WEGNER[1], RYAN YAMACHIKA[1], JEREMY E. DAHL[2], ROBERT M. K. CARLSON[2], STEVEN G. LOUIE[1] and MICHAEL F. CROMMIE[1]*

[1]Department of Physics, University of California at Berkeley, and Materials Sciences Division, Lawrence Berkeley Laboratory, Berkeley, California 94720-7300
[2]MolecularDiamond Technology, Chevron Technology Ventures, Richmond, California 94802
*e-mail: yywang@berkeley.edu, crommie@berkeley.edu

**Diamondoids are a unique form of carbon nanostructure best described as hydrogen-terminated diamond molecules[1]. Their diamond-cage structures and tetrahedral *sp³* hybrid bonding create new possibilities for tuning electronic band gaps, optical properties, thermal transport, and mechanical strength at the nanoscale[1,2]. The recently-discovered higher diamondoids[3,4] (each containing more than three diamond cells) have thus generated much excitement in regards to their potential versatility as nanoscale devices[5-15]. Despite this excitement, however, very little is known about the properties of isolated diamondoids on metal surfaces, a very relevant system for molecular electronics. Here we report the first molecular scale study of individual tetramantane diamondoids on Au(111) using scanning tunneling microscopy and spectroscopy. We find that both the diamondoid electronic structure and electron-vibrational coupling exhibit unique spatial distributions characterized by pronounced line nodes across the molecular surfaces. *Ab-initio* pseudopotential density functional calculations reveal that the observed dominant electronic and vibronic properties of diamondoids are determined by surface hydrogen terminations, a feature having important implications for designing diamondoid-based molecular devices.**

The [121]tetramantane diamondoid studied here consists of four diamond cages face-fused into a straight rod with a chemical formula $C_{22}H_{28}$ (see schematic structure in Fig. 1a)[3]. The 28 surface dangling carbon bonds are all saturated by hydrogen atoms. The [121]tetramantanes used here were extracted from petroleum and isolated and purified to a white crystalline powder with purity greater than 99% by weight. Purification procedures included distillation, thermal processing, and both size- and shape-selective high performance liquid chromatography, as described in Ref. 4.

Tetramantane molecules were thermally evaporated onto a clean Au(111) substrate held at room temperature in ultrahigh vacuum. Under these conditions the diamondoid molecules self-assemble into an ordered overlayer. Fig. 1b shows the scanning tunneling microscope (STM) topography taken at $T = 7$ K of a sub-monolayer of tetramantane on Au(111) where the oval-shaped tetramantane molecules form a close-packed structure with lattice constants 11.4 Å × 8.3 Å (see red arrows), and an apparent height of ~3.3 Å. The formation of long-range ordering at room temperature implies high molecular mobility and weak bonding between the diamondoids and the Au(111) substrate.

In order to examine individual tetramantane diamondoids, we used the STM tip to manipulate the diamondoids from the edge of the island onto an empty Au(111) terrace at $T = 7$ K. Fig. 1c and d display STM topographs of the same individual diamondoid taken with different sample biases. The image taken at sample bias $V = +2$ volts (Fig. 1c), which probes the unoccupied local electronic density of states (LDOS), exhibits pronounced line nodes across the molecular surface. In contrast, the image taken at -2 volts (Fig. 1d) has less pronounced features, revealing a much weaker spatial dependence of the occupied electronic states. Individual tetramantane was found to have a number of different orientations on Au(111). Fig. 2a shows STM images of three different tetramantane diamondoids lying on Au(111) with different molecular orientations, as seen by the distinctive spatial patterns of the line nodes. Diamondoids can be switched between molecular orientations by dragging and rotating the tetramantane using the STM tip.

The local electronic structure of tetramantane diamondoid was probed by scanning tunneling spectroscopy (STS). Fig. 3a shows the *dI/dV* spectra of the bare Au(111) substrate (black curve), and the spatially-averaged *dI/dV* of a tetramantane diamondoid residing on Au(111) (blue curve). The main features of the two spectra are very similar, implying that the diamondoid makes a small contribution to the total surface electronic LDOS. Subtracting the Au background from the total *dI/dV* yields the contribution of tetramantane to the spectrum, which is shown by the red curve in Fig. 3a. The molecular electronic LDOS is quite low for the energy range from -2.5 eV to +2.5 eV, consistent with a highly insulating ground state having an energy gap $\Delta \geq 5$ eV about the Fermi level ($E_F$)[9,14]. Although no sharp resonances related to the highest occupied molecular orbital (HOMO) or the lowest unoccupied molecular orbital (LUMO) were observed, the gentle increase of *dI/dV* above $E_F$ may represent the tail of the broadened LUMO level.

High-resolution *dI/dV* spectroscopy in the low bias range reveals a "V"-like feature about $E_F$ and two steep jumps in *dI/dV* at ±356 mV (the upper panel of Fig. 3b). The numerical derivative of this curve ($d^2I/dV^2$) correspondingly shows two sharp anti-symmetric peaks at ±356 mV (the lower panel of Fig. 3b). This is the classic hallmark of inelastic electron tunneling spectroscopy (IETS), in which the electronic excitation of molecular vibrations opens a new tunneling channel[16]. The 356 meV vibrational energy seen here for tetramantane corresponds to the stretch mode of a C-H bond, as has been observed in numerous other hydrocarbons[17]. The "V"-like feature in the *dI/dV* between the ±356 mV jumps might also be related to inelastic tunneling of electrons interacting with quasi-continuous, low-frequency vibrations of the tetramantane[18].

In order to understand the microscopic characteristics of electron-vibration coupling across the surface of a single diamondoid, we performed spatially resolved IETS across individual tetramantane molecules[19-21]. We find that the pronounced inelastic signals at ±356 mV only exist on certain parts of the molecules, and become negligibly small elsewhere. Fig. 4a shows the $d^2I/dV^2$ maps of the C-H stretch mode for three individual diamondoids having different orientations, demonstrating unambiguously that the vibronic coupling strength is strongly localized to narrow slices on the molecular surfaces. Comparing the IETS maps of the diamondoids with their respective topographs (Fig. 4b) reveals an anti-correlation between the elastic and inelastic tunneling channels: the IETS intensity peaks sharply at the topographic line nodes where elastic tunneling is suppressed.

Such localization of the inelastic signal is unexpected because the 28 C-H bonds surrounding the tetramantane present a dense and nearly uniform surface termination. Moreover, one might naively expect that enhanced elastic tunneling goes hand-in-hand with enhanced inelastic tunneling (since there is then more electron density to excite molecular vibrations)[22], opposite to the anti-correlation observed here.

In order to understand this behavior, we have carried out *ab-initio* pseudopotential density functional theory (DFT) calculations within the local density approximation (LDA) for a [121]tetramantane molecule on Au(111). We performed these calculations first by using a plane-wave pseudopotential code to obtain the properties of the isolated molecule[23]. We then modeled the gold surface by three layers of 56 gold atoms each (in a supercell geometry) and calculated the properties of the combined diamondoid/Au(111) system using the SIESTA code[24], which employs a localized basis set, since the use of plane waves for the combined calculation would be much more expensive.

Fig. 3c depicts the schematic energy diagram of the system under investigation. The theoretical Kohn-Sham HOMO-LUMO gap of the isolated molecule is found to be 5.2 eV, which is an under-estimation of the true quasiparticle gap[25]. We have calculated an electron affinity/ionization potential gap of 7.9 eV for tetramantane by computing the total energy of singly charged molecules. When a diamondoid is placed on a Au(111) surface, however, screening from the metal substrate and hybridization with the surface states can significantly alter these properties. Fig. 3d displays the DFT-calculated electronic density of states of a tetramantane molecule when the effect of the Au(111) surface is included. The LDA HOMO and LUMO levels are located at approximately -1.0 eV and +3.9 eV relative to $E_F$, leading to a reduced Kohn-Sham energy gap of 4.9 eV. The broadened LDOS of the LUMO level extends into the energy gap as a shallow tail, consistent with the experimental *dI/dV* curve. However, the theoretical HOMO resonance at -1.0 eV is not observed in the experiment down to -2.5 eV.

This discrepancy can be understood by calculating the spatial distribution of the wavefunctions of the molecular electronic states. Fig. 2d shows the calculated isosurfaces of the HOMO and LUMO electronic wavefunction square at a value corresponding to 50% of the charge of each state for an isolated tetramantane with a particular molecular orientation. The occupied electronic states are concentrated on the centers of the C-C bonds, reflecting the spatial localization of the *sp³* bonding orbitals. Such confined states are difficult for STM to detect (with the tip at several Ångstroms above the molecule), which explains the weak spatial features of the negative-bias image and the absence of the HOMO resonance in the *dI/dV* spectrum. The calculated LUMO orbital, on the other hand, is much more delocalized in space and exhibits pronounced spatial variations similar to those seen experimentally.

Closer comparison of the tetramantane LUMO isosurface and the diamondoid molecular structure uncovers an intriguing trend: the strong spatial variation of the electronic states is closely related to the nature of the surface hydrogen terminations. The large, delocalized LUMO state-density exists only at the 12 singly-hydrogenated CH sites, and forms smooth patches in these CH-rich areas. The 8 doubly-hydrogenated $CH_2$ sites (highlighted by dashed red circles in Fig. 2), in contrast, exhibit little electronic density and form depressions in the LUMO isosurface plot. This is caused by the peculiar formation of the LUMO orbitals on the $CH_2$ sites. Here the most relevant LUMO wavefunction, the so called $\sigma^*_{CH2}$ orbital, is constructed by a linear combination of the anti-bonding wavefunctions formed between the two canted carbon $sp^3$ orbitals and between the two hydrogen $s$ orbitals[26]. The derived $\sigma^*_{CH2}$ wavefunction has a pronounced node on the mirror plane that bisects the H-C-H complex.

This behavior can be seen better in Fig. 2b which shows three calculated LUMO-level STM topographs of isolated tetramantanes. The schematic topviews in Fig. 2c reveal that the three molecules lie on the substrate with their *z*-axis orientated along the diamond crystallographic [111], [110], and [100] directions, respectively (from left to right). These simulated images match well the essential features of the corresponding STM images taken at +2.0 volts (Fig. 2a). The excellent agreement clearly demonstrates that the pronounced nodal features found in the tetramantane STM images result from suppressed LUMO electronic state-density at the doubly-hydrogenated $CH_2$ sites.

We next turn to the inelastic tunneling spectroscopy. The strength of electronic coupling to the C-H stretch mode has been evaluated by calculating the change of the electronic energy eigenvalues with respect to displacements along the canonical phonon coordinates[21, 22]. Out of the 28 C-H stretch modes we have identified three that interact strongly with tunneling electrons within 15 meV of the experimental signal. The question remains, however, as to why the experimental IETS signal is enhanced in the $CH_2$-terminated regions of the diamondoid and suppressed elsewhere. We postulate that this occurs because the $CH_2$ region has a denser concentration of C-H bonds, and hence a higher probability of electronic interactions with the C-H stretch mode compared to the singly-hydrogenated CH region. Unique local vibrational features at the $CH_2$ sites may also enhance the strength of electron-vibrational coupling in these regions.

Our unified STM and DFT investigations show that the important electronic and vibronic properties of higher diamondoids are determined by microscopic differences in the surface hydrogen termination where the singly-hydrogenated CH sites and doubly-hydrogenated $CH_2$ sites behave very differently. The pronounced nodal (anti-nodal) features in STM topography (IETS) originate exclusively from the $CH_2$ sites. This information should help to structurally optimize the effect of C-H stretch vibrations on electron transport in future molecular devices. Our results also suggest that selective substitution of H atoms on the singly or doubly hydrogenated sites may lead to very different functionalities for electronic devices[15].

## Methods

### Sample preparation and STM experiments

In our experiments, tetramantane molecules were thermally deposited in ultrahigh vacuum (UHV) from a Knudsen cell evaporator onto a clean Au(111) surface held at room temperature. STM measurements were performed using a homebuilt system with a polycrystalline PtIr tip operated at a cryogenic temperature $T = 7$ K. STM topography was carried out in a constant current mode, and *dI/dV* spectra and images were measured through lock-in detection of the ac tunneling current driven by a 450 Hz, 1-10 mV (rms) signal added to the junction bias (defined as the sample potential referenced to the tip) under open-loop conditions. The typical parameters for STM molecular manipulation are $V = 5$ mV and $I = 1$ nA.

### Theoretical calculations

Our calculations were performed using density functional theory (DFT) in the local density approximation (LDA) using the functional of Perdew and Zunger[27]. We employed the plane-wave pseudopotential method (PW-PP)[23] using Troullier-Martins norm-conserving pseudopotentials[28]. We used a 60 Ry plane-wave cutoff and a rectangular simulation cell of 40×40×50 $a_B$ ($a_B$ is the Bohr radius) For our SIESTA calculations[24] we used the same pseudopotentials as in our PW-PP calculations and a double-ζ polarized basis. We extended the basis set by introducing ghost atoms in the vacuum region and by including two s-type orbitals per ghost atom.

## Acknowledgments

This work was supported in part by NSF Grant Nos. DMR04-39768, EEC-0425914, COINS, UC Discovery grant ELE 05-10234. Computational resources have been provided by DOE at the National Energy Research Scientific Computing Center. Y.W. thanks the Miller Institute for a research fellowship. E. K. is a fellow of the Onassis Foundation. D.W acknowledges support by the Alexander von Humboldt Foundation.

**Figure Captions:**

**Figure 1 STM topography of tetramantane molecules. a,** The structural model of a [121]tetramantane ($C_{22}H_{28}$). **b,** Constant-current topograph of a sub-monolayer of tetramantane on Au(111) taken with sample bias $V = 2.0$ volts and tunneling current $I = 50$ pA. **c** and **d,** STM topographs (25 Å × 25 Å) of an individual tetramantane taken with sample bias +2.0 V and -2.0 V, respectively.

**Figure 2 STM images and DFT simulations of individual tetramantane. a,** Positive-bias STM images of three individual tetramantanes with different molecular orientations. **b,** DFT-simulated LUMO-level STM topographs of the three isolated tetramantanes. **c,** Topview of the schematic structure of the three tetramantanes lying on the substrate with their *z*-axis orientated along the diamond crystallographic [111], [110], and [100] directions, respectively. Dashed red circles highlight the doubly-hydrogenated $CH_2$ sites. **d,** DFT-calculated isosurfaces of the HOMO and LUMO electronic wavefunction square at a value corresponding to 50% of the charge of the state for the third molecule shown in Fig. 2a.

**Figure 3 *dI/dV* spectroscopy of tetramantane molecules. a,** *dI/dV* spectra of bare Au(111) (black curve) and a tetramantane molecule on Au (blue curve). The difference (red curve) represents the contribution of tetramantane. **b,** Low-bias *dI/dV* spectrum on a tetramantane shows two steps at ±356 mV (upper panel). The numerically derived $d^2I/dV^2$ curve shows two anti-symmetric peaks at ±356 mV (lower panel). **c,** Schematic electronic energy diagram of tetramantane. **d,** The calculated density of states of a tetramantane molecule on Au(111) compared with the experimental *dI/dV* spectrum.

**Figure 4 Spatial maps of the IETS intensity. a,** $d^2I/dV^2$ maps of the C-H stretch mode (0.36 volt) of three tetramantane molecules with different orientations. The inelastic signal is strongly localized to narrow slivers on the molecular surface. **b,** STM topographs of the same molecules shown in **a** reveal that the inelastic intensity peaks sharply at the LUMO-level topographic line nodes.

**a** [121]tetramantane, $C_{22}H_{28}$

- C
- H

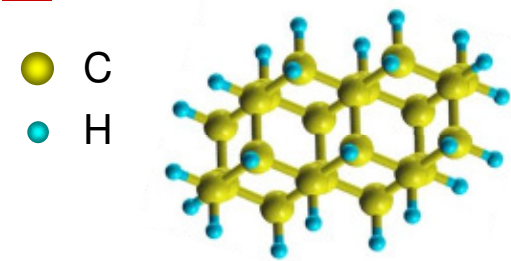

Bias-dependent topographs

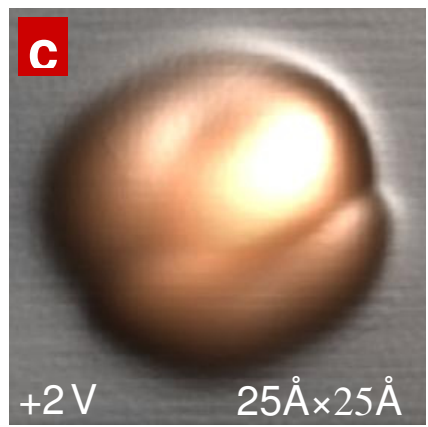

c

+2 V    25Å×25Å

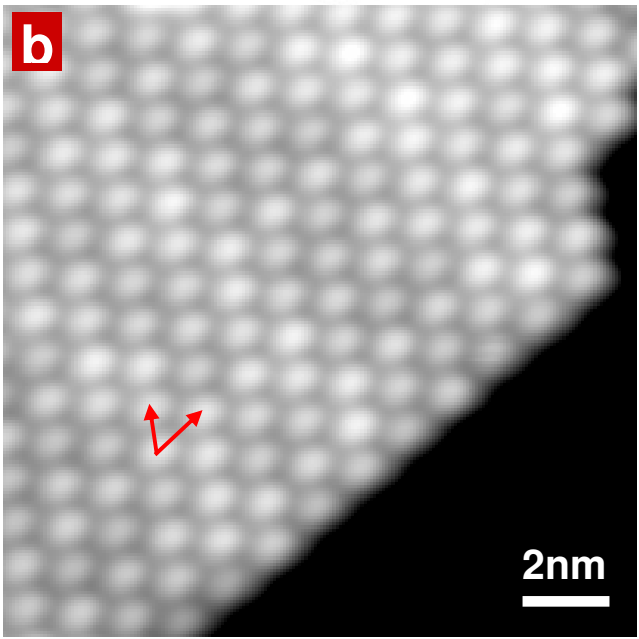

b

2nm

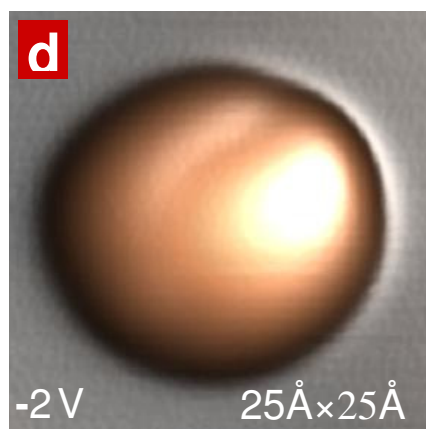

d

−2 V    25Å×25Å

Figure 1

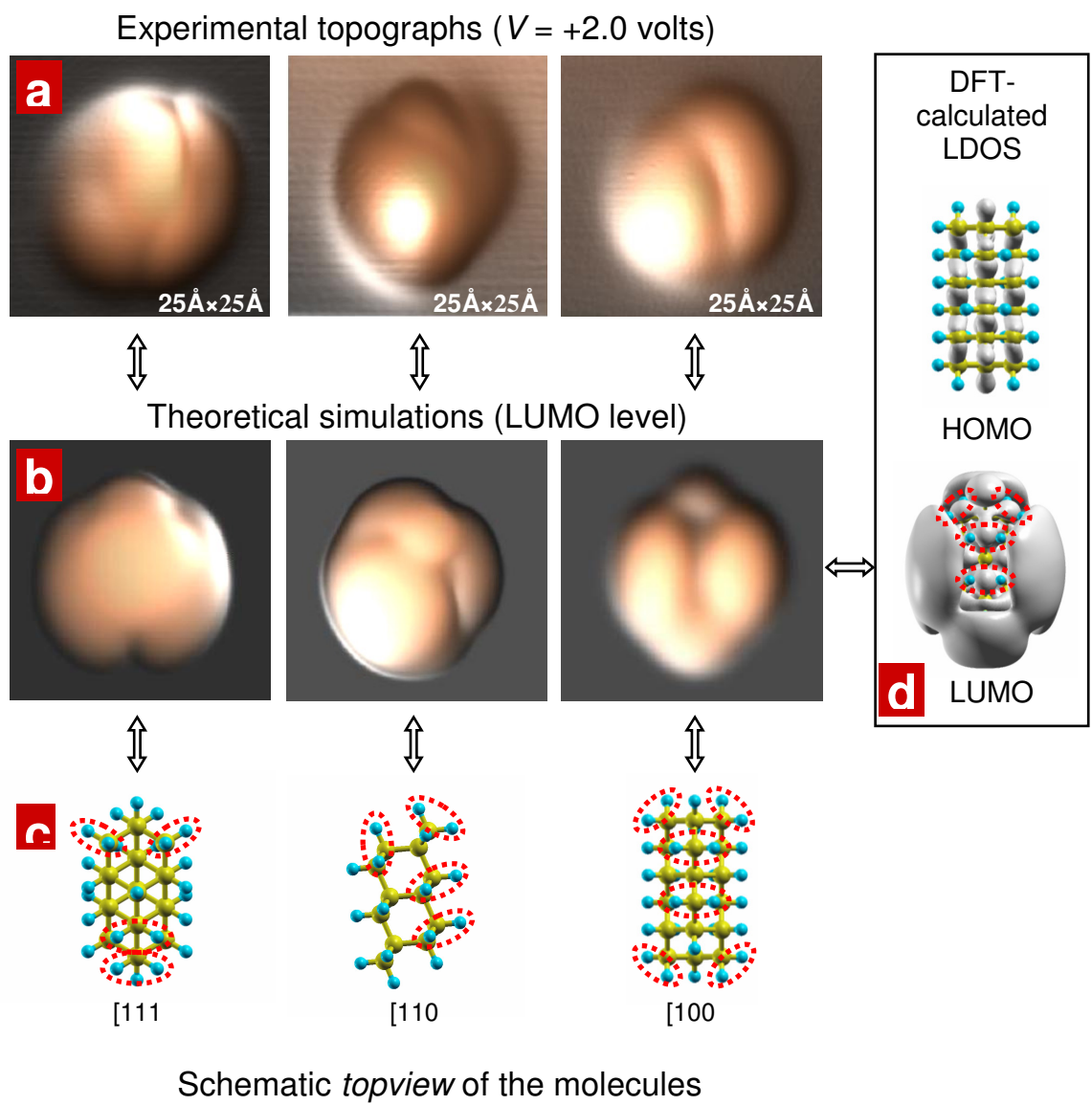

Figure 2

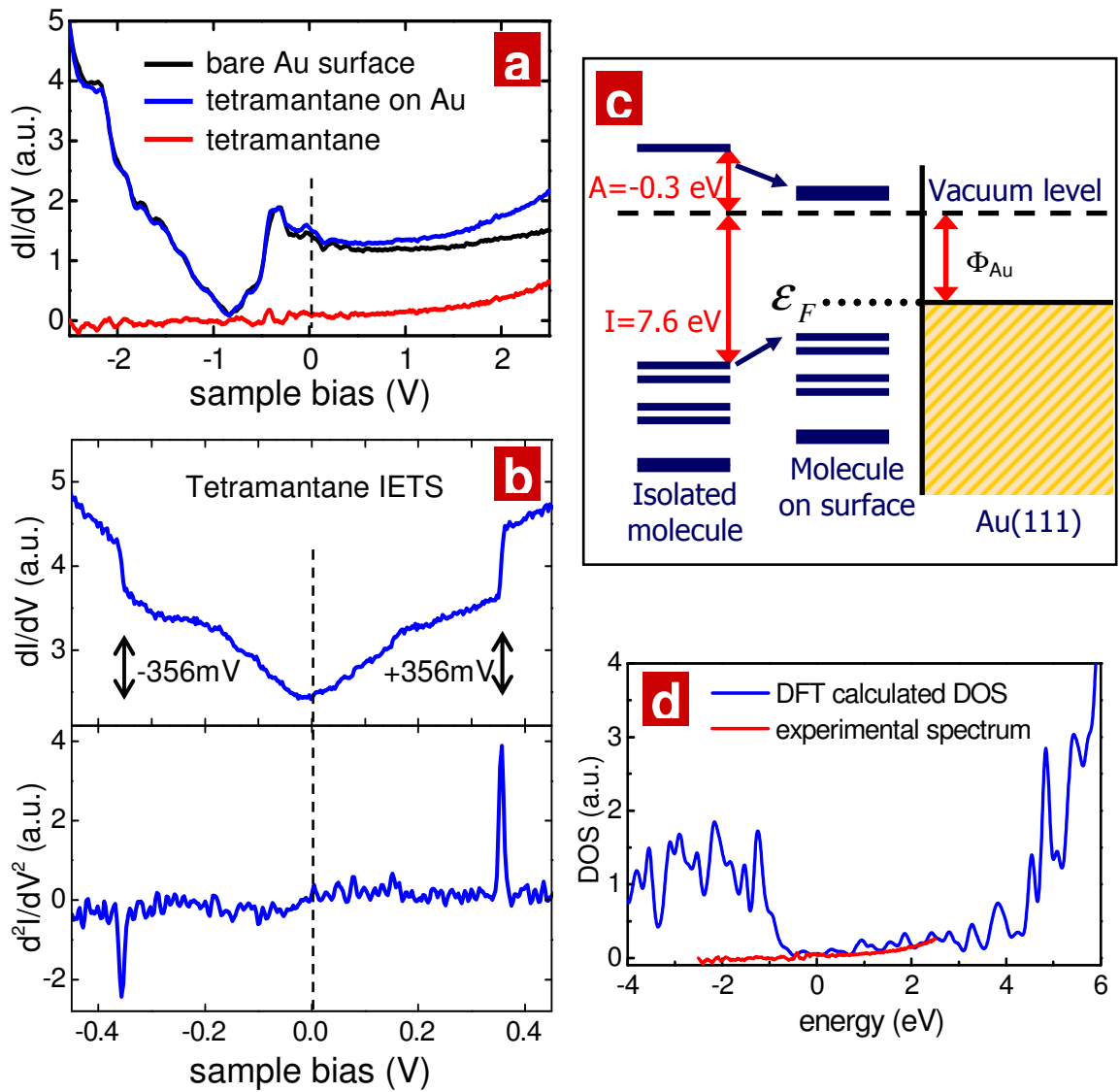

Figure 3

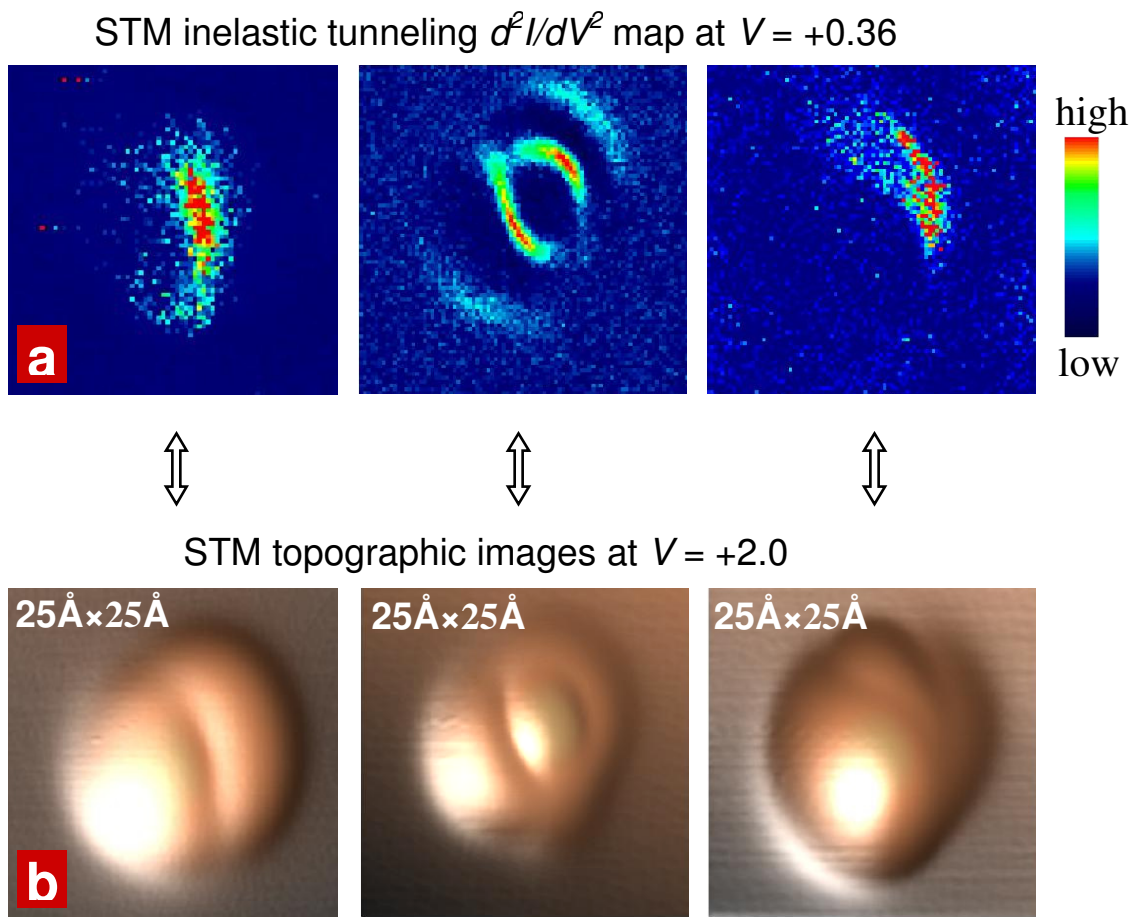

Figure 4